\documentclass{kapproc}
\usepackage{psfig}
\kluwerbib
\begin{document}

\articletitle[Empirical diagnostics of HII regions based on sulphur lines]{EMPIRICAL DIAGNOSTICS OF HII REGIONS BASED ON SULPHUR EMISSION LINES}

\author{Enrique P\'{e}rez Montero}

\email{enrique.perez@uam.es}

\author{\'Angeles I. D\'{\i}az}
\email{angeles.diaz@uam.es}

\affil{C-XI. Departamento de F\'{\i}sica Te\'orica. \\
Universidad Aut\'{o}noma de Madrid. 28049. Cantoblanco. Madrid}

\begin{abstract}

Spectrophotometry constitutes a unique way to obtain the whole 
diagnostics (density, temperature, ionic abundances) of ionized gas
nebulae, thus providing invaluable information about the objects where
they reside. If such nebulae are in the low excitation regime, the 
diagnostics have to relie on the observation of the more intense
emission lines. We have compiled data for a 
sample of HII regions, GEHR and HII galaxies and compared them with 
results derived from photoionizacion models whose parameters cover 
the physical conditions of the nebulae. Our results confirm a lower 
uncertainty for the diagnostics when using empirical methods based on
the strong lines of [SII] and [SIII] replacing and complementing those 
based on the [OII] and [OIII] lines.
\end{abstract}

\section*{Introduction}

The most accurate method to ascertain chemical abundances in io-nized
gas nebulae relies on the measurement of the auroral lines and the 
determination of the electron line temperatures. Since the nature of these
lines is collisional, in low excitation nebulae, where the electron 
temperature is low as well, it is often difficult  to detect them in the
optical. There are many studies in the literature that 
propose and analyze new empirical calibrators, based on strong emission 
lines in the optical ([OII], [OIII], [NII]), where the auroral lines 
are too weak. The flux of these lines is governed, as the inner 
ionization structure of the HII regions, by the metallicity of the gas, the 
effective temperature of the ionizing radiation and the ionization 
parameter, $U$ (i.e. the rate between Lyman ionizing photons and 
the density of particles) which takes into account geometrical
effects. The goal of the our models  is to 
isolate the dependence of the lines on the metal content of the nebula.

In this work, we have compiled a sample of diffuse HII regions in 
the Galaxy,  giant extragalactic HII regions and HII galaxies, all of
them with a direct 
determination of the oxygen abundance and measurements of the 
strong lines of [OII], [OIII], [NII], [SII] and [SIII] in the 
optical and near-infrared. We have compared 
these data with the results from photoionization models 
(Cloudy: Ferland, 1996) under realistic physical conditions using as
ionization spectra those provided by Co-Star stellar model atmospheres
and we have studied the dependence 
on the three functional parameters of the strong line intensities. 

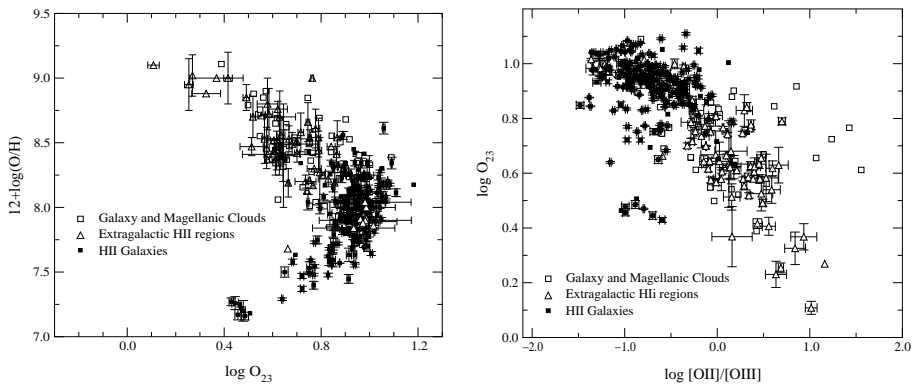
\vspace*{-0.3cm}

\begin{center}
\begin{figure}[bh]
\sidebyside
{\psfig{figure=perez-montero-fn1.eps,height=5cm,width=5.8cm,angle=-90,clip=}}
{\psfig{figure=perez-montero-fn2.eps,height=5cm,width=5.8cm,angle=-90,clip=}}
\end{figure}
\end{center}
\vspace*{-0.5cm}

%\vspace*{-0.5cm}
\begin{center}
\begin{figure}[b]
\sidebyside
{\psfig{figure=perez-montero-fn3.eps,height=5cm,width=5.8cm,angle=-90,clip=}}
{\psfig{figure=perez-montero-fn4.eps,height=5cm,width=5.8cm,angle=-90,clip=}}
\sidebyside
{\psfig{figure=perez-montero-fn5.eps,height=5cm,width=5.8cm,angle=-90,clip=}}
{\psfig{figure=perez-montero-fn6.eps,height=5cm,width=5.8cm,angle=-90,clip=}}
\end{figure}
\end{center}

\vspace{-1.8cm}

\section{The O$_{23}$ (or R$_{23}$) parameter}

It is defined as the sum of the lines [OII]$\lambda$3727{\AA}, 
[OIII]$\lambda$$\lambda$4959,5007{\AA}{\AA} relative to H$\beta$. The main features 
of this parameter are its two-valued nature (see upper figure, 
at left) and its dependence on ionization parameter (at right, 
the ratio [OII]/[OIII] depends on log $U^{-1}$) . At high metallicities,
the low electron temperatures cause the strong oxygen lines to 
decrease, and, on the contrary, at low metallicities the increasing 
oxygen abundance cause the line fluxes to be stronger.  We have therefore 
an upper branch of the relation, for 12+log(O/H) $>$ 8.4, where almost no
direct determinations of abundances exist and that has to
be calibrated with the help of theoretical models.  Our
photoionization models predict an uncertainty of 0.2 dex in the oxygen
abundance derived in this regime. In the lower 
branch, for 12+log(O/H) $<$ 8.0, whose dependence on ionization 
parameter can be considered explicitely, our models 
predict 0.15 dex of uncertainty. The uncertainty associated to the
oxygen abundance determination in  the intermediate regime 
8.0 $<$ 12+log(O/H) $<$ 8.4, where the calibration reverses,
can be high however much larger than this ($\approx$ 0.5-0.6
dex. Unfortunately,  about 40\% 
of the objets (and about 80\% of the HII galaxy sample) lie on this zone.

\section{The N2 parameter}

This parameter is defined as the ratio of $\lambda$[NII]6584{\AA} and H$\alpha$. 
In the figure below (top left) it can be seen that the parameter 
is single-valued for all metallicities, but Denicol\'o et al. (2001) point 
out its strong dependence on ionization parameter (top right) and the 
uncertainty due to N/O variations. 
Nevertheless N2 represent an improvement for objects located in the
turn over region of the oxygen abundance {\em vs } O$_{23}$
relation. Our models predict an uncertainty  of 0.3 dex associated to
the abundances determined through this parameter at any metallicity. 

\section{The S$_{23}$ parameter}

Defined as the sum of the strong lines of [SII]$\lambda$$\lambda$6717,6731{\AA}{\AA} and 
[SIII] $\lambda$$\lambda$9069,9532{\AA}{\AA} and used as a metallicity calibrator by
D\'{\i}az \& P\'erez-Montero (DPM00: 2000). It has a lower dependence on
ionization parameter as can 
be seen below (bottom right) and it remains single-valued until 
metallicity about 1.5 
times solar (12+log(O/H)$_\odot$ = 8.69). Both facts are
confirmed by our models, from which and using the compiled set of data
together with those in DPM00 we deduce the relation: 
$$12+log(O/H) = 8.29 + 2.02\log S_{23} + 0.72(\log S_{23})^2$$
with an uncertainty of 0.15 dex. Some authors (Oey \&
Shields, 2000) have suggested a modified
parameter S$_{234}$ that would include he contribution of the
[SIV]$\lambda$10.5$\mu$ line. This contribution would be relevant only for high
excitation objects and, since very few data exist at present,
it could be calculated only from photo-ionization models. Therefore 
the uncertainty would not be decreased.

\section{The S$_{23}$/O$_{23}$ parameter}

Finally, we think that this parameter is very promissing to provide  a 
consistent abundance calibration over the whole range of metallicities 
and therefore. In fact,  it can be very useful to describe large
trends, such as the behaviour of abundance gradients in 
spiral galaxies. Our models predict a rather strong dependence on ionization 
parameter (due to O$_{23}$) but  a preliminar calibration ca be given 
(below, left):
\begin{equation}
\nonumber
12+\log(O/H) = 9.2 + 0.8\cdot \log \left(\frac{S_{23}}{O_{23}}\right) - 0.6 \cdot  \log ^2 \left(\frac{S_{23}}{O_{23}}\right)
\end{equation}
that applied to the disk of spirals, such as M101 
(Garnett et al., 1997) shows the exponential nature of the abundance
distribution (below, right).

\vspace*{-0.3cm}

\begin{center}
\begin{figure}[bh]
\sidebyside
{\psfig{figure=perez-montero-fn7.eps,height=4.9cm,width=5.8cm,angle=-90,clip=}}
{\psfig{figure=perez-montero-fn8.eps,height=4.9cm,width=5.8cm,angle=-90,clip=}}
\end{figure}
\end{center}

\end{document}